\documentclass[aps,prl,twocolumn,showpacs,citeautoscript,superscriptaddress]{revtex4-1}

\usepackage{graphicx}  
\usepackage{dcolumn}   
\usepackage{bm}        
\usepackage{amssymb}   
\usepackage[none]{hyphenat}  
\usepackage{nicefrac}  
\usepackage{amsmath}   
\usepackage{footmisc}  
\usepackage{latexsym}  
\usepackage{romannum}  
\usepackage{hyperref}  
\usepackage{multirow}  

\renewcommand{\arraystretch}{1.5}

\newcommand{\kv}{\mathbf{k}}

\newcommand{\kp}{k_{\parallel}}

\renewcommand{\Im}{\mathrm{Im}}

\begin{document}

\title{Nonequilibrium Casimir effects of nonreciprocal surface waves}

\author{Chinmay Khandekar} \email{ckhandek@purdue.edu}
\affiliation{Department of Electrical Engineering, Stanford
  University, California 94305, USA}

\author{Siddharth Buddhiraju} 
\affiliation{Department of Electrical Engineering, Stanford
  University, California 94305, USA}

\author{Paul R. Wilkinson} \affiliation{Department of Chemistry and
  Biochemistry, University of California Los Angeles, Los Angeles,
  California 90095, USA}

\author{James K. Gimzewski}
\affiliation{Department of Chemistry and
  Biochemistry, University of California Los Angeles, Los Angeles,
  California 90095, USA}

\author{Alejandro W. Rodriguez} \affiliation{Department of Electrical
  and Computer Engineering, Princeton University, Princeton, New
  Jersey 08544, United States}

\author{Charles Chase} \affiliation{UnLAB, Savannah, Georgia 31405, USA}

\author{Shanhui Fan} \email{shanhui@stanford.edu}
\affiliation{Department of Electrical Engineering, Stanford
  University, California 94305, USA}

\date{\today}

\begin{abstract}
  We show that an isotropic dipolar particle in the vicinity of a
  substrate made of nonreciprocal plasmonic materials can experience a
  lateral Casimir force and torque when the particle's temperature
  differs from that of the slab and the environment. We connect the
  existence of the lateral force to the asymmetric dispersion of
  nonreciprocal surface polaritons and the existence of the lateral
  torque to the spin-momentum locking of such surface waves. Using the
  formalism of fluctuational electrodynamics, we show that the
  features of lateral force and torque should be experimentally
  observable using a substrate of doped Indium Antimonide (InSb)
  placed in an external magnetic field, and for a variety of
  dielectric particles. Interestingly, we also find that the
  directions of the lateral force and the torque depend on the
  constituent materials of the particles, which suggests a sorting
  mechanism based on lateral nonequilibrium Casimir physics.
\end{abstract}

 \maketitle

Nonreciprocal electromagnetic surface waves can occur at an interface
between two semi-infinite bulk regions, at least one of which breaks
reciprocity. Notable examples include surface waves at the interface
between gyrotropic medium and regular dielectric as well as one-way
edge modes at the interface between electromagnetic analogues of
topological and regular
insulators~\cite{camley1987nonreciprocal,yu2008oneway,
  haldane2008possible,wang2009observation,wang2008reflection,
  Zhang2020broadband}. In many cases, the existence of such
nonreciprocal surface states is linked with the nontrivial topological
behavior of the bulk region making them fundamentally
intriguing~\cite{Gangaraj2019unidirectional,lu2014topological,
  silveirinha2016bulk}. They are also useful for practical devices
like isolators, circulators, phase shifters and
lasers~\cite{hu2015surface,bahari2017nonreciprocal}.

\begin{figure}[t!]
  \centering\includegraphics[width=0.8\linewidth]{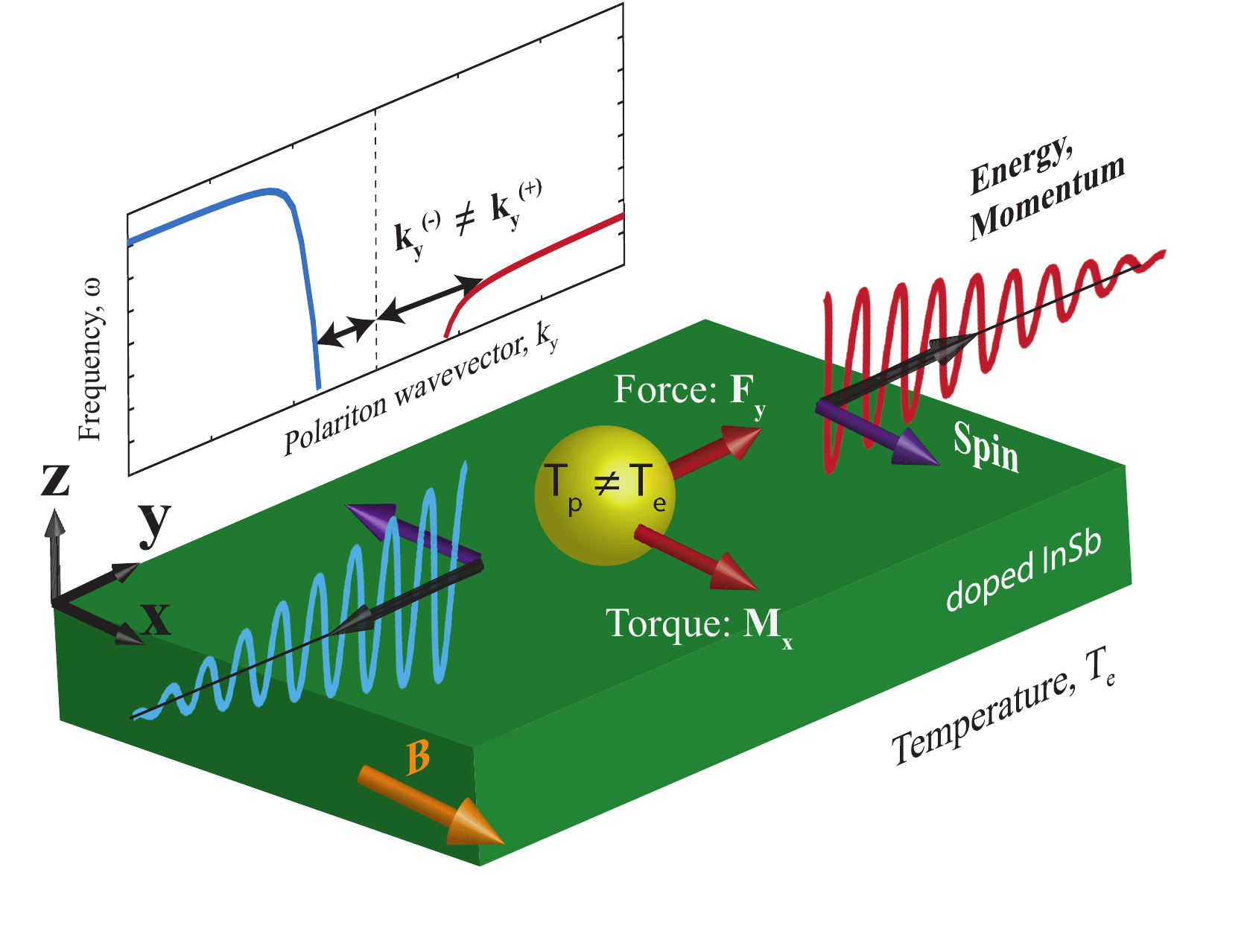}
  \caption{We illustrate nonequilibrium lateral Casimir force and
    torque on an isotropic dipolar nanoparticle in the near field of a
    doped InSb slab placed in magnetic field parallel to its
    surface. We assume $B=1$T is applied along $\mathbf{x}$ axis of
    the geometry and clarify the asymmetric interaction of the
    particle with nonreciprocal SPPs of the slab along $\mathbf{y}$
    axis (Voigt configuration). The asymmetric dispersion shown in the
    inset reveals the unequal magnitude of linear momentum for forward
    and backward SPPs of equal frequency. The SPPs also carry
    transverse spin angular momentum locked to their momentum as
    displayed. When the particle's temperature $T_p$ is different from
    the temperature $T_e$ of the rest of the system, the net exchange
    of linear and spin angular momentum via fluctuations-generated
    photons is asymmetric with respect to forward and backward SPPs
    causing a lateral force transverse to applied field ($F_y$) and a
    lateral torque parallel to applied field ($M_x$).}
  \label{scm1}
\end{figure}

In this letter, we show that the nonreciprocal surface waves have
intriguing fundamental implications for Casimir
physics~\cite{gong2020recent, bimonte2017nonequilibrium}. As
illustrated in Fig.~\ref{scm1}, we consider an exemplary system
comprising a reciprocal nanoparticle in the vicinity of a gyrotropic
plasmonic substrate, consisting of a doped InSb slab in the presence
of a magnetic field, $B=1$T, applied parallel to its surface. The slab
supports nonreciprocal surface plasmon polaritons (SPPs) whose
asymmetric dispersion ($\omega(\kv)\neq \omega(-\kv)$) is depicted in
the inset under a Voigt configuration. These SPPs carry not only
linear momentum but also spin angular momentum locked transverse to
the linear momentum~\cite{van2016universal,bliokh2015quantum}. When
the particle's temperature ($T_p$) differs from the temperature of its
surroundings including the slab ($T_e$), there is a net exchange of
quantum- and thermal-fluctuations-generated photons between the
particle and the nonreciprocal SPPs. Since the resulting exchange of
linear momentum and spin angular momentum is asymmetric with respect
to forward and backward SPPs, the particle experiences a lateral
nonequilibrium Casimir force ($F_y$) transverse to applied magnetic
field, and a lateral nonequilibrium Casimir torque ($M_x$) parallel to
magnetic field.

We provide a rigorous fluctuational electrodynamic analysis of these
effects applicable to a particle in the vicinity of a bianisotropic
substrate, not limited to the Voigt configuration of the
schematic. Instead of the usual trace formulas in Casimir physics, we
provide semi-analytic expressions for force and torque which are more
transparent in capturing the flow of energy, linear and angular
momenta between the particle and the SPPs. Using these expressions, we
prove that a lateral force and torque on an isotropic particle are not
possible for reciprocal media, or for either reciprocal or
nonreciprocal media under equilibrium condition ($T_p =
T_e$). Moreover, since the particle interacts with large-wavevector
SPPs where nonlocal dielectric response cannot be ignored, we use a
nonlocal hydrodynamic magnetoplasma model of InSb which indicates the
absence of strictly unidirectional
SPPs~\cite{buddhiraju2020absence}. Our calculations show that the
lateral force and torque arise because of nonreciprocity, and do not
require strictly unidirectional states.

Fluctuational phenomena in nonreciprocal systems is an emerging
research topic with recent theoretical proposals for new fundamental
effects~\cite{zhu2016persistent,ben2016photon,khandekar2019thermal,
  guo2020theoretical} and practical
applications~\cite{zhao2020axion,ott2019radiative,moncada2020magnetic,
  khandekar2020new} related to radiative heat transfer. Recent works
on this topic explored the effects of nonreciprocity on Casimir torque
and force for closely separated
plates~\cite{Lindel2018rotation,gelbwaser2020near} and torque for
individual sphere~\cite{guo2020single} or closely-separated
cubes~\cite{gao2021thermal}. However, the important fundamental
connection between the properties of nonreciprocal surface waves and
the behaviors of nonequilibrium Casimir forces and torques has not
been discussed previously. Moreover, experiments probing these effects
are challenging in plate-plate
geometry~\cite{Lindel2018rotation,gelbwaser2020near} because they are
quite weak due to the large inertia of bulk plates. In contrast, our
concrete predictions for a sphere-plate geometry indicate that these
effects can be readily detected in modern
Casimir~\cite{gong2020recent,bimonte2017nonequilibrium} and nanoscale
heat transfer~\cite{narayanaswamy2008near,kim2015radiative}
experiments. We also reveal interesting dependence of directionalities
of the lateral features on the material composition of the particle.
We note that the lateral Casimir forces for equilibrium systems have
been studied previously in
Ref.~\cite{Chen2002lateral,Emig2003normal,Bao2018inhomogeneity,
  Manjavacas2017lateral,Muller2016anisotropic}. In these systems,
lateral force can only arise from nonuniform lateral configurations.
In contrast, lateral nonequilibrium Casimir force in our nonreciprocal
system can arise despite the translational invariance of the substrate
in the lateral directions.



\emph{Theory:} We consider a dipolar particle characterized by
isotropic polarizability tensor $\alpha_{jk}=\alpha_0
\mathrm{I}_{3\times 3}$ and placed at a distance $d$ from a generic
bianisotropic planar substrate. The particle is at temperature $T_p$
and the entire environment surrounding the particle, including the
substrate and the vacuum half-space, is at temperature $T_e$.  Using
fluctuational electrodynamics (see supplement for full derivation), we
obtain the following semi-analytic expressions for the spectral
density of net power transfer from the particle to the environment
($P$) as well as force ($F_j$) and torque ($M_j$) for $j\in [x,y,z]$
on the particle:
\begin{widetext}
  \begin{align}
    \label{Pw}
P(\omega) &= (\Theta_{T_p}-\Theta_{T_e})k_0^2 \Im\{\alpha_0(\omega)\}
\bigg[\frac{k_0}{\pi} + \int_{0}^{\infty} d\kp \int_0^{2\pi} d\phi
  \Im\bigg(\frac{i\kp e^{2ik_zd}}{8\pi^2
    k_z}[r_{ss}+r_{pp}(2\kp^2/k_0^2-1)]\bigg)\bigg] \\ \label{Fxw}
F_x(\omega) &= -(\Theta_{T_p}-\Theta_{T_e}) \frac{k_0}{c}
\Im\{\alpha_0(\omega)\} \int_{0}^{\infty} d\kp \int_0^{2\pi} d\phi
\Im\bigg( \frac{i\kp e^{2ik_zd}}{8\pi^2
  k_z}[r_{ss}+r_{pp}(2\kp^2/k_0^2-1)] \kp\cos\phi \bigg)
\\ F_z(\omega) &= -(\Theta_{T_p}+\Theta_{T_e}) \frac{k_0}{c}
\Im\{\alpha_0(\omega)\} \int_{0}^{\infty} d\kp \int_0^{2\pi} d\phi
\Im\bigg(\frac{i\kp e^{2ik_zd}}{8\pi^2
}[r_{ss}+r_{pp}(2\kp^2/k_0^2-1)] \bigg) \\ \label{Mxw} M_x(\omega) &=
-(\Theta_{T_p}-\Theta_{T_e})\frac{k_0}{c} \Im\{\alpha_0(\omega)\}
\int_{0}^{\infty} d\kp \int_0^{2\pi} d\phi \frac{\kp^2}{8\pi^2 k_0}
\bigg[\cos\phi \Im\{\frac{(r_{sp}-r_{ps})e^{2ik_z d}}{k_z}\} +
  2\sin\phi \Im\{\frac{r_{pp}e^{2ik_zd}}{k_0} \} \bigg] \\ M_z(\omega)
&= (\Theta_{T_p}-\Theta_{T_e})\frac{k_0}{c} \Im\{\alpha_0(\omega)\}
\int_{0}^{\infty} d\kp \int_0^{2\pi} d\phi \frac{\kp}{8\pi^2 k_0} \Im[
  (r_{sp}+r_{ps})e^{2ik_z d}] \\ F_y(\omega) &= F_x(\omega,\cos\phi
\rightarrow \sin\phi), \hspace{20pt} M_y(\omega)=M_x(\omega,\cos\phi
\rightarrow \sin\phi, \sin\phi \rightarrow -\cos\phi) \label{Fyw}
\end{align}
\end{widetext}
Here, $\Theta_T(\omega,T)=\hbar\omega/2 +
\hbar\omega/[\mathrm{exp}(\hbar\omega/k_B T)-1]$ is the mean energy of
a harmonic oscillator of frequency $\omega$ at thermodynamic
temperature $T$. Total power, force, torque are obtained by
integrating corresponding spectral density over all frequencies as
$Q^t = \int_{-\infty}^{\infty} Q(\omega) d\omega/2\pi$ where
$Q=\{P,F_j,M_j\}$. The term outside the wavevector integration in
Eq.~\ref{Pw} comes from the heat transfer between the particle and the
vacuum half-space. Such transfer does not lead to any force or torque
for an isotropic particle. The terms inside the wavevector integration
indicate the interaction between the particle and the substrate via
quantum- and thermal-fluctuations-generated photons characterized by
their frequency $\omega$ and in-plane wavevector components
$(k_x=\kp\cos\phi,k_y=\kp\sin\phi)$ where $\phi$ is the angle with
$\mathbf{x}$ axis of the geometry. $k_z$ is the perpendicular
wavevector component such that $k_z^2 + \kp^2 = k_0^2$ where
$k_0=\omega/c$. $r_{jk}$ for $j,k \in [s,p]$ denotes the Fresnel
reflection coefficient which is the amplitude of $j$-polarized
reflected light under incident unit-amplitude $k$-polarized light
characterized by $(\omega,\kp,\phi)$. Electromagnetic response of the
substrate enters through the reflection coefficients. Here we assume
that there are no relative translational or rotational motions between
the particle and the substrate. In the presence of such relative
motions, Eqs.(\ref{Pw})-(\ref{Fyw}) will need to be modified to take
into account resulting Doppler shift in the frequency of photons being
exchanged~\cite{pendry2010quantum,leonhardt2010comment,
  volokitin2011quantum,guo2020single}. The extension of
Eqs.(\ref{Pw})-(\ref{Fyw}) taking into account the rotation of the
particle, is provided in the supplement. We also note that the
correction due to such relative motions are in general quite small
since these motions are usually in the non-relativistic regime. Thus,
in practice, Eqs.(\ref{Pw})-(\ref{Fyw}) should be sufficient for most
experimental situations.

We note that all components of force and torque can be separately
derived by dividing the integrand in the power spectral density
[Eq.~\ref{Pw}] by the photon energy $\hbar\omega$, multiplying by
linear momentum vector $\hbar(\kp\cos\phi,\kp\sin\phi,kz)$ and spin
angular momentum vector $\hbar(\sin\phi,-\cos\phi,0)$ of polaritons,
and accounting for whether the photon is emitted or absorbed by the
particle. This physically meaningful alternative derivation is exact
for force calculation but approximate for torque calculation because
the spin angular momentum $\hbar(\sin\phi,-\cos\phi,0)$ is exact only
for large wavevector polaritons ($\kp \gg k_0$) [also proved
  separately in the supplement]. The alternative derivation reveals
the intimate connection between energy, linear and angular momentum
transfer in our system.

Equations \ref{Fxw}-\ref{Fyw} are consistent with some of the general
physics constraints. At thermal equilibrium ($\Theta_{T_p} =
\Theta_{T_e}$), all force and torque components as predicted by
Eqs.~\ref{Fxw}-\ref{Fyw} are zero except the perpendicular force
$F_z$. This follows from the observation that the Casimir free energy
of the system depends on the vertical displacement $d$ between the
particle and the substrate but not the lateral displacement.  For a
substrate made of reciprocal media, the Fresnel coefficients satisfy
the reciprocity constraints
$r_{ss,pp}(\kp,\phi)=r_{ss,pp}(\kp,\phi+\pi)$ and
$r_{sp}(\kp,\phi)=-r_{ps}(\kp,\phi+\pi)$~\cite{bimonte2007general,
  khandekar2020new}. By integrating Eqs.~\ref{Fxw} to \ref{Fyw}, it
follows that there is no lateral force and no lateral torque on the
particle in the near field of a reciprocal medium, for both
equilibrium and nonequilibrium scenarios. For a nonreciprocal medium,
the reciprocity constraints no longer hold and therefore, one can
expect to see nontrivial effects for the nonequilibrium
scenarios. Therefore, we consider a substrate made of doped InSb with
external magnetic field which results in a strong nonreciprocity in
its electromagnetic response. The reflection coefficients in
equations~\ref{Pw} to \ref{Fyw} are obtained by solving the boundary
conditions at the interface~\cite{buddhiraju2020absence}.

\begin{figure}[t!]
  \centering\includegraphics[width=0.95\linewidth]{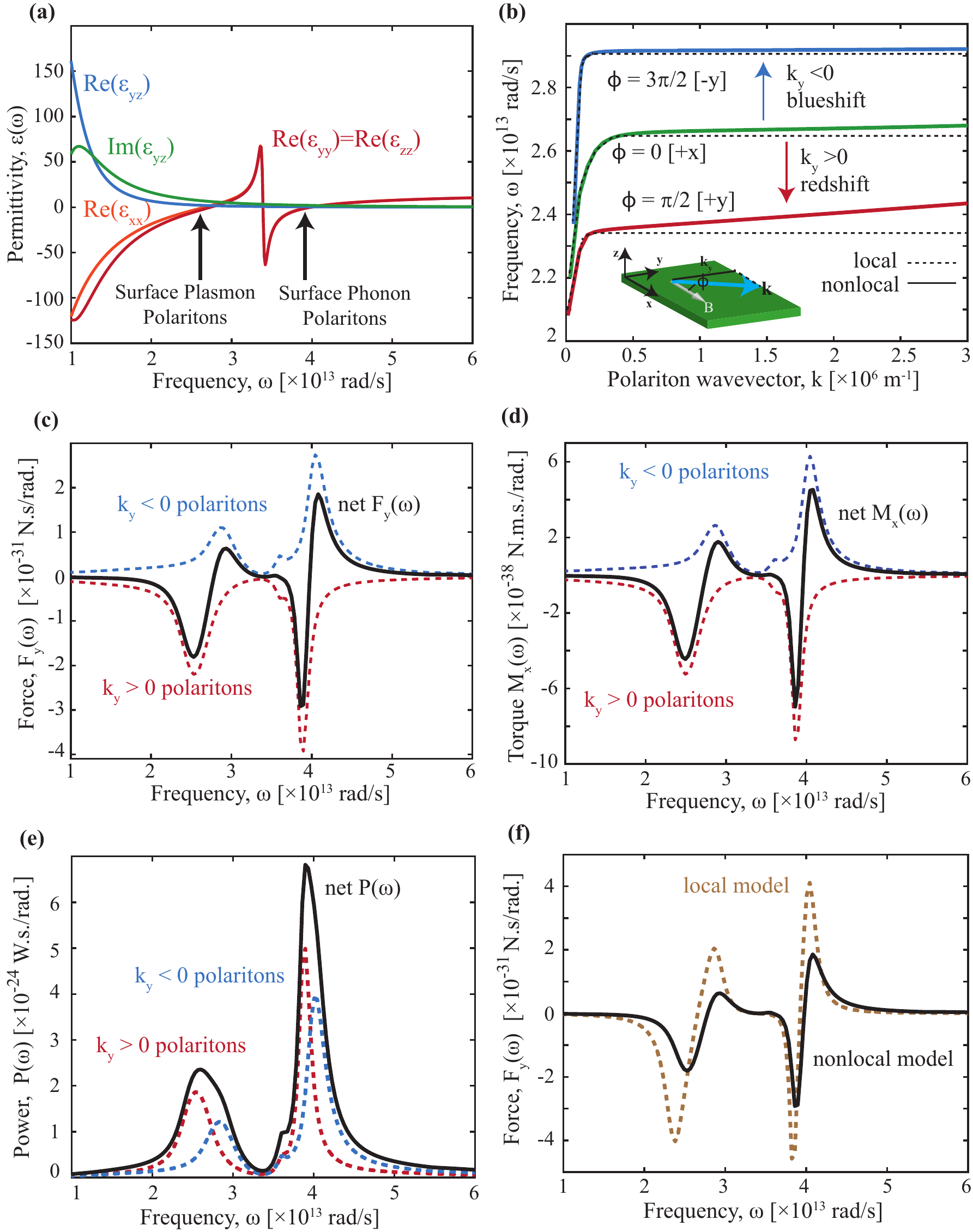}
  \caption{(a) The frequency-dependent local permittivity of InSb slab
    in magnetic field $B=1$T reveals the approximate locations of
    surface polaritons. (b) Dispersion $\omega(\kp)$ for SPPs
    propagating at an in-plane angle $\phi$ with $B$ field shows
    redshift for $\phi \in [0,\pi)$ ($k_y > 0$) and blueshift for
      $\phi \in [\pi, 2\pi)$ ($k_y < 0$). Dispersion is plotted using
        both a nonlocal model (solid lines) and a local model (dashed
        lines). (c) The force spectral density calculated using a
        nonlocal model for a particle at a distance $d=0.3\mu$m from
        the slab and under operating temperatures $T_p=305$K,
        $T_e=300$K is shown by black line. Its characteristic shape
        follows from the collective contributions of red-shifted ($k_y
        >0$) forward polaritons and blue-shifted ($k_y < 0$) backward
        polaritons which are plotted separately as dashed lines. (d)
        and (e) respectively demonstrate the power and torque spectral
        densities (black lines). The separately plotted collective
        contributions of red-shifted and blue-shifted polaritons
        (dashed lines) clarify the asymmetric flow of energy and
        angular momenta respectively. (f) shows the force spectral
        density obtained using local and nonlocal models.}
  \label{fig1}
\end{figure}

\emph{Results:} We consider both a local model and a nonlocal
model~\cite{buddhiraju2020absence} for the permittivity of InSb to
highlight some of the consequences of the nonlocal dielectric
response. Figure \ref{fig1}(a) shows the frequency-dependent local
permittivity ($\epsilon_{jk}(\omega)$ for $j,k=[x,y,z]$) of InSb slab
in presence of magnetic field $B=1T$ applied along $\mathbf{x}$ axis.
The variation of diagonal components reveals the approximate locations
of the surface phonon (higher branch) and plasmon (lower branch)
polaritons, henceforth denoted by the acronym SPhP and SPP,
respectively. Fig.~\ref{fig1}(b) shows the polaritonic dispersion
($\omega(\kp)$) of nonreciprocal SPPs propagating in different
directions. Since the behavior is the same for nonreciprocal SPhPs, we
focus on SPPs in this figure. As shown in the inset, because of the
Zeeman interaction between the spin magnetic moment of nonreciprocal
surface polaritons and the magnetic field~\cite{khandekar2019thermal},
the polaritons making an angle $\phi \in [0,\pi)$ with B-field and
  carrying positive momentum $k_y > 0$ experience a redshift whereas
  the polaritons characterized by $\phi\in[\pi, 2\pi)$ or $k_y < 0$
    experience a blueshift compared to the dispersion in the absence
    of the magnetic field ($B=0$). Based on the momentum $k_y$, the
    red-shifted polaritons can be collectively interpreted as forward
    waves and blue-shifted polaritons as backward waves. Due to these
    opposite frequency shifts, the forward and backward momentum waves
    of equal frequency carry very different momentum and experience
    unequal near-field coupling with the particle (proportional to
    $e^{-2|k_z|d}$ in Eqs.~\ref{Pw}-\ref{Fyw}), leading to a lateral
    Casimir force and torque.
    
Figure.~\ref{fig1}(b) shows the SPP dispersion using both a nonlocal
model (solid lines) and a local model (dashed lines). At large
wavevectors, $\omega(\kp)$ increases based on a nonlocal model while
it reaches a constant value based on a local model. While this reveals
a stark contrast between the two models pertaining to presence or
absence of strictly unidirectional SPPs~\cite{buddhiraju2020absence},
we note that the lateral force and torque arise due to opposite
frequency shifts of forward and backward waves noted above, which
occurs for both local and nonlocal models and does not require strict
unidirectionality. Nonetheless, the magnitudes of force and torque are
slightly different as we show further below. Unless noted otherwise,
all results below are generated using a nonlocal model.

To highlight the effects of the nonreciprocal surface waves, we first
calculate the spectral densities in Eqs.~\ref{Fxw}-\ref{Fyw} assuming
a frequency-independent polarizability,
$\Im\{\alpha_0(\omega)\}=10^{-19}$m$^{-3}$ for the particle. We use
the temperatures $T_p=305$K and $T_e=300$K. We find that
$F_x(\omega)$, $M_y(\omega)$, $M_z(\omega)$ are identically zero for
this configuration. We do not discuss the perpendicular Casimir force
$F_z(\omega)$ here since its behavior is well known, and focus instead
on the lateral effects. Figure~\ref{fig1}(c) demonstrates the force
spectral density (black line) when the center of the particle is at a
distance $d=0.3\mu$m from the surface along with the separate
collective contributions of red-shifted polaritons (red dashed line)
and blue-shifted polaritons (blue dashed line). In the case of $T_p >
T_e$, there is a net emission from the particle to the substrate.  The
emission to red-shifted and blue-shifted polaritons provides a
negative $(F_y < 0)$ and positive $(F_y > 0)$ contribution to the
force, from linear momentum conservation, as they carry positive $(k_y
> 0)$ and negative $(k_y < 0)$ momentum, respectively. The difference
in these two contributions arising from the nonreciprocity in the
dispersion of surface waves, results in the lateral Casimir
force. Fig.~\ref{fig1}(d) demonstrates the torque spectral density
(black line) where the sign of the torque follows from the separate
contributions of red-shifted and blue-shifted polaritons carrying
positive and negative transverse spin (along $\mathbf{x}$ axis)
respectively, and based on angular momentum conservation. And again,
the difference in these two collective contributions causes the
lateral Casimir torque.

Figure \ref{fig1}(e) also highlights the asymmetry in the net power
transfer from the particle to forward and backward nonreciprocal
surface waves and the resulting spectrally broadened total
contribution (black line). Figure \ref{fig1}(f) plots the force
spectral density based on local and nonlocal models. The local model
overestimates the magnitude of the force and its spectrum is
red-shifted compared to that obtained using the nonlocal model.
  
\begin{figure}[t!]
  \centering\includegraphics[width=0.95\linewidth]{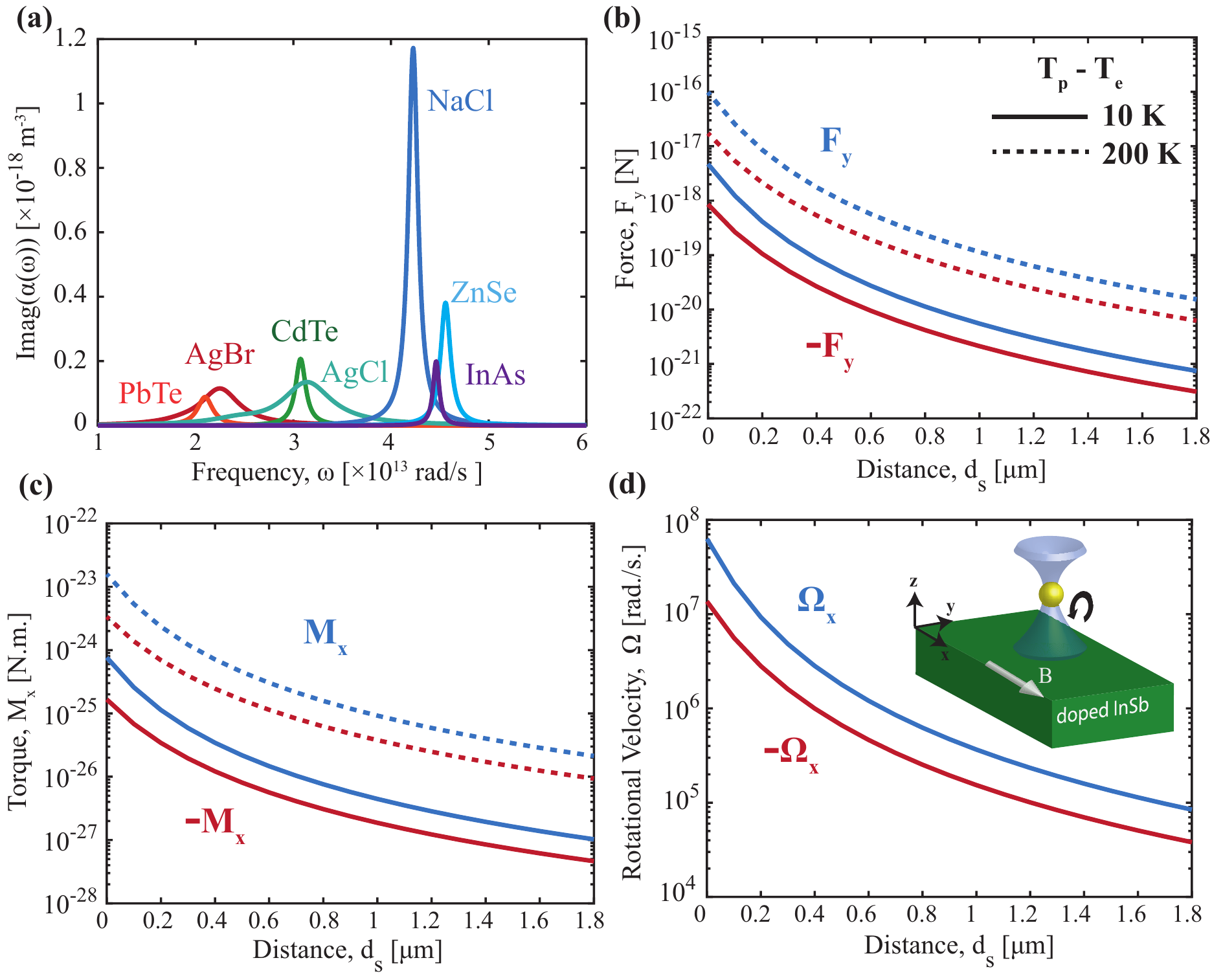}
  \caption{(a) We consider nanoparticles of radius $R=200$nm and made
    of different materials such that they exhibit dipolar resonances
    over the frequency range of interest. (b,c) demonstrate the
    dependence of force and torque on surface-to-surface distance
    $d_s=d-R$ for AgBr (red) and NaCl (blue) nanoparticles for two
    different operating temperatures $T_p=310$K (solid lines) and
    $T_p=500$K (dashed lines) assuming $T_e=300$K. (d) We consider a
    potential experiment where the same AgBr and NaCl particles are
    trapped near InSb slab in $B=1T\mathbf{x}$ inside a vacuum chamber
    of pressure $10^{-5}$torr. The particles reach a steady state
    angular velocity at which the Casimir torque is balanced by the
    rotational drag of imperfect vacuum. The figure plots the distance
    dependence of steady-state rotational velocities for operating
    temperatures $T_p=310$K and $T_e=300$K.}
  \label{fig2}
\end{figure}

To estimate the magnitude of the total lateral force and torque in
potential experiments, we consider various nanoparticles (PbTe, AgBr,
CdTe, AgCl, NaCl, ZnSe, undoped InAs) of radius $R=200$nm whose
polarizability response functions are shown in Fig.\ref{fig2}(a). The
material permittivity $\epsilon(\omega)$ is obtained from various
references~\cite{palik1998handbook,foteinopoulou2019phonon,caldwell2015low}
and the polarizability is calculated as $\alpha(\omega)=4\pi R^3
[\epsilon(\omega)-1]/[\epsilon(\omega)+2]$ using a dipolar
approximation which is fairly accurate since the particle radius is
much smaller than the relevant wavelengths of $45$-$90\mu$m (deep
subwavelength regime). The stated materials were chosen because they
exhibit a strong resonant response in the aforementioned frequency
range. By comparing the resonant frequency of the particle with the
frequencies of red-shifted and blue-shifted nonreciprocal SPPs and
SPhPs of InSb, we can deduce the directionalities of the lateral force
and torque. Assuming warm particles ($T_p > T_e$), it follows that
AgBr, PbTe particles experience $F_y^t<0$ and $M_x^t<0$ since they
emit photons dominantly to red-shifted SPPs while all other particles
experience $F_y^t>0$ and $M_x^t>0$ since they emit photons dominantly
to blue-shifted SPPs or SPhPs. The opposite force behavior for
different materials suggests that it might be possible to use such
lateral Casimir force to sort particles based on the material
composition.

Figure~\ref{fig2}(b,c) demonstrate the distance dependence of lateral
force and torque on AgBr (red lines) and NaCl (blue lines) particles
of radius $R=200$nm for temperatures $T_p=310$K (solid lines) and
$T_p=500$K (dashed lines) assuming $T_e=300$K. The magnitudes of force
and torque increase as the particle's temperature increases. For both
choices of $T_p$, their magnitudes increase as the particle-surface
separation decreases. For NaCl particle at $T_p=500$K, lateral force
$F_y^t \sim 10^{-16}$N and lateral torque $M_x^t \sim 10^{-23}$Nm are
observed when the vacuum gap spacing between the particle and the
substrate is reduced to few nanometers. Such a force is in fact
comparable to the force of gravity on the particle. For NaCl particle
of mass density $\rho=2.165g/$cm$^3$, gravity force is $W=7.3\times
10^{-17}$N. For AgBr particle of mass density $\rho=6.47g/$cm$^3$,
gravity force is $W=2.17\times 10^{-16}$N. Ultrasensitive detection of
force of magnitude $10^{-20}$N and torque of magnitude $10^{-27}$Nm
have been realized in
experiments~\cite{gong2020recent,bimonte2017nonequilibrium,
  antognozzi2016direct,gieseler2013thermal}. Other recent experiments
have measured near-field heat transfer at nanoscale under large
temperature
differentials~\cite{bhatt2020integrated,narayanaswamy2008near,
  kim2015radiative,desutter2019near,ghashami2018precision}. In
Ref.~\cite{kim2015radiative}, temperature differences larger than
$100$K were maintained between a tip and a plate separated by
$2$nm. Therefore, we are optimistic that the experimental
demonstration of our predictions should be feasible.

We also consider an alternative experiment where the particle is
optically levitated in the near field of InSb slab inside a vacuum
chamber. Due to the Casimir torque, the particle experiences angular
acceleration. As its angular velocity increases, it experiences an
oppositely directed rotational drag torque due to residual gas
molecules of the imperfect vacuum. Eventually, such a particle will
reach a steady-state angular velocity at which the nonequilibrium
Casimir torque is balanced by the rotational drag torque. Assuming a
vacuum chamber of gas pressure $10^{-5}$ torr, Fig.~\ref{fig2}(d)
shows the steady state rotational velocities for levitated AgBr (red)
and NaCl (blue) particles under temperature difference of
$T_p-T_e=10$K (see supplement for more details). As displayed, the
particles rotate with opposite angular velocities and they reach
rotational velocities in the MHz-GHz range depending on their distance
from the surface. Such rotations could be detected in experiments. We
note that the nonreciprocity-induced nonequilibrium torque was
recently analyzed for a single gyrotropic
particle~\cite{guo2020single} and for two finite-size gyrotropic
particles separated by small distance~\cite{gao2021thermal}. The
magnitude of torque in our work is few orders of magnitude larger than
the values reported in the previous studies due to the large density
of surface polariton states in the near field of a gyrotropic
susbtrate~\cite{xu10enhancement}. We also note that the magnitudes of
the lateral nonequilibrium Casimir force and torque can be further
enhanced using Weyl semimetals~\cite{kotov2018giant} which provide
much stronger gyrotropy compared to InSb considered here.


\emph{Conclusion:} We demonstrated that nonreciprocal surface waves
can lead to a nonequilibrium lateral Casimir force and torque on an
isotropic nanoparticle. We clarified the origin of these effects by
transparently accounting for the underlying asymmetric flow of energy
and momenta. We connected the lateral force to the dispersion of
nonreciprocal surface polaritons and the lateral torque to the
spin-momentum locking. We also made predictions for potential
experiments to detect these nonreciprocal Casimir effects soon. Our
work indicates intriguing opportunities at the intersection of
nonreciprocity, photon spin, Casimir physics and topological
materials.

\emph{Acknowledgements:} This work is supported by the DARPA Defense
Science Office (Grant No. HR00112090080). 

\bibliography{photon}
 
\end{document}


\title{Nonequilibrium Casimir effects of nonreciprocal surface waves:
  Supplementary Materials}

\author{Chinmay Khandekar} \email{ckhandek@purdue.edu}
\affiliation{Department of Electrical Engineering, Stanford
  University, California 94305, USA}

\author{Siddharth Buddhiraju} 
\affiliation{Department of Electrical Engineering, Stanford
  University, California 94305, USA}

\author{Paul R. Wilkinson} \affiliation{Department of Chemistry and
  Biochemistry, University of California Los Angeles, Los Angeles,
  California 90095, USA}

\author{James K. Gimzewski}
\affiliation{Department of Chemistry and
  Biochemistry, University of California Los Angeles, Los Angeles,
  California 90095, USA}

\author{Alejandro W. Rodriguez} \affiliation{Department of Electrical
  and Computer Engineering, Princeton University, Princeton, New
  Jersey 08544, United States}

\author{Charles Chase} \affiliation{UnLAB, Savannah, Georgia 31405, USA}

\author{Shanhui Fan} \email{shanhui@stanford.edu}
\affiliation{Department of Electrical Engineering, Stanford
  University, California 94305, USA}

\date{\today}

\begin{abstract}
We derive the semi-analytic expressions for power, force and torque
spectral densities provided in the main text using fluctuational
electrodynamics. For making predictions for potential experiments, we
rigorously analyze the rotational dynamics of the nanoparticle in a
vacuum chamber, taking into account an additional rotational friction
arising from the residual air molecules. While the Casimir torque is
derived assuming a non-rotating particle, we also prove that this
derivation is accurate as long as the particle's rotational angular
velocity ($\Omega$) is much smaller than the relevant thermal
fluctuations frequencies ($\omega \sim 10^{13}$rad/s). We separately
derive and plot the angular-velocity-dependent Casimir torque.
\end{abstract}

\pacs{}
\maketitle

\onecolumngrid

\section{Fluctuational Electrodynamic derivation of power, force and torque}

We consider a dipolar spherical particle of polarizability
$\alpha_{jk}=\alpha_0 \mathcal{I}_{3\times 3}$ placed at a distance
$d$ from the semi-infinite half-space of a generic linear,
time-invariant bianisotropic medium.  The particle is at thermodynamic
temperature $T_p$ and the rest of the system, denoted as the
environment is at temperature $T_e$. Using fluctuational
electrodynamics, the spectral densities of the net power transfer from
the particle to the environment $P(\omega)$, the force $F_j(\omega)$
and the torque $M_j(\omega)$ on the particle for $j=[x,y,z]$
are~\cite{henkel2002radiation,Manjavacas2017lateral}:
\begin{align}
P(\omega) &= \omega \Im[\langle {p}_j^{\text{fl}^*}(\omega)
  E_j^{\text{ind}}(\omega) \rangle + \langle
  p_j^{\text{ind}^*}(\omega) E_j^{\text{fl}}(\omega)
  \rangle] \label{Pexp} \\ F_j(\omega) &= \Re[ \langle
  p_k^{\text{fl}}(\omega)\partial_j E_k^{\text{ind}^*}(\omega)\rangle
  + \langle p_k^{\text{ind}}(\omega)\partial_j
  E_k^{\text{fl}^*}(\omega) \rangle ] \label{Fexp} \\ M_j(\omega) &=
\varepsilon_{jkl} \Re[\langle
  p_k^{\text{fl}}(\omega)E_l^{\text{ind}^*}(\omega) \rangle + \langle
  p_k^{\text{ind}}(\omega)E_l^{\text{fl}^*}(\omega)
  \rangle] \label{Mexp}
\end{align}
where Einstein summation convetion is used with $j,k,l \in [x,y,z]$.
$\langle \cdots \rangle$ denotes statstical ensemble average of the
fluctuating quantities enclosed within. These physical quantities are
expressed in SI units and the total power, force, torque are obtained
by integrating over both positive and negative frequencies as $Q^t =
\int_{-\infty}^{\infty} Q(\omega) d\omega/2\pi$ where
$Q=\{P,F_j,M_j\}$. Here
$p_j^{\text{ind}}=\epsilon_0\alpha_{jk}E_k^{\text{fl}}$ is the dipole
moment induced in the particle by the environment field fluctuations.
$E_j^{\text{ind}}$ is the field at the dipole location by the
fluctuating dipole moment itself. In general, the field
$E_j^{\text{ind}}(\rv_2,\omega)$ at an arbitrary location $\rv_2$
generated by the dipole moment $p_j^{\text{fl}}(\rv_1,\omega)$ at the
spatial position $\rv_1$ is given by
$E_j^{\text{ind}}(\rv_2,\omega)=\omega^2\mu_0G_{jk}(\rv_2,\rv_1,\omega)
p_j^{\text{fl}}(\rv_1,\omega)$ where $G_{jk}(\rv_2,\rv_1,\omega)$ is
the Green's function. The semi-analytic form of Green's function in
the dipole-plate geometry considered here
is~\cite{khandekar2019thermal}:
\begin{align}
\overline{\overline{G}}(\rv_1,\rv_2,\omega) &=\int \frac{d^2
  \mathbf{k}_\parallel}{(2\pi)^2} e^{i\mathbf{k}_\parallel \cdot
  (\mathbf{R}_1-\mathbf{R}_2)} \frac{i}{2k_z} \bigg[
  \overbrace{e^{ik_z(z_1-z_2)}[\ev_{s+}\ev_{s+}^T +
      \ev_{p+}\ev_{p+}^T]}^{\text{vacuum part}} \nonumber \\ &\hspace{60pt} +
  \underbrace{e^{ik_z(z_1+z_2)}[(r_{ss}\ev_{s+}+r_{ps}\ev_{p+})\ev_{s-}^T
      + (r_{sp}\ev_{s+}+r_{pp}\ev_{p+})\ev_{p-}^T]}_{\text{reflected
      part}} \bigg] \label{Green}
\end{align}
where $\rv = (\mathbf{R},z)$ is the position
vector. $\kv_\parallel=(k_\parallel \cos\phi, k_\parallel \sin\phi)$
is the in-plane momentum such that $\phi$ is the angle with
$\mathbf{x}$ axis of the geometry. $k_z$ is the perpendicular
wavevector component such that $k_z^2 + \kp^2 = k_0^2$ where
$k_0=\omega/c$. $r_{jk}$ for $j,k \in [s,p]$ denotes the Fresnel
reflection coefficient which is the amplitude of $j$-polarized
reflected light when unit-amplitude $k$-polarized light characterized
by $(\omega,\kp,\phi)$ is incident.  The polarization vectors
$\ev_{j\pm}$ for $j=[s,p]$ for waves going along $\pm \ev_z$ direction
are:
\begin{align}
\ev_{s\pm}=\begin{bmatrix} \sin\phi \\ -\cos\phi
\\ 0 \end{bmatrix}, \hspace{30pt} \ev_{p\pm} =
\frac{-1}{k_0} \begin{bmatrix} \pm k_z\cos\phi \\ \pm k_z \sin\phi
  \\ -k_\parallel \end{bmatrix}
\end{align}

The calculation of the power, force, torque in
Eq.~\ref{Pexp}-\ref{Mexp} relies on $p_j^{\text{fl}}$ and
$E_j^{\text{fl}}$ which denote statistically uncorrelated particle
dipole moment fluctuations and environment field fluctuations at the
dipole location respectively. They satisfy the following fluctuation
dissipation theorems (FDTs) based on linear response
theory~\cite{landau2013course}:
\begin{align}
  \langle p_j^{\text{fl}}(\omega) p_k^{\text{fl}^*}(\omega') \rangle
  &= \frac{\alpha_{jk}(\omega) - \alpha_{kj}^*(\omega)}{2i}
  \frac{\epsilon_0}{\omega}\Theta_{T_p}
  (2\pi)\delta(\omega-\omega') \label{pfdt} \\ \langle
  E_j^{\text{fl}}(\rv_1,\omega)E_k^{\text{fl}^*}(\rv_2,\omega') \rangle
  &= \frac{G_{jk}(\rv_1,\rv_2,\omega)-
    G_{kj}^*(\rv_2,\rv_1,\omega)}{2i}\mu_0\omega\Theta_{T_e}
  (2\pi)\delta(\omega-\omega') \label{Efdt}
\end{align}
where $\langle \cdots \rangle$ denotes the statistical ensemble
average and $\Theta_T(\omega,T)=\hbar\omega/2 +
\hbar\omega/[\mathrm{exp}(\hbar\omega/k_B T)-1]$ is the Planck's
function giving the mean energy of a harmonic oscillator of frequency
$\omega$ at thermodynamic temperature $T$. In the derivation of the
above spectra [Eqs.\ref{Pexp},\ref{Fexp},\ref{Mexp}] from the real-valued
dipole moment $p_j(t)=\int_{-\infty}^{\infty} \frac{d\omega}{2\pi}
p_j(\omega)e^{-i\omega t}$ and electric field
$E_j(t)=\int_{-\infty}^{\infty} \frac{d\omega}{2\pi} E_j e^{-i\omega
  t}$, the factor of $(2\pi)$ in the above FDTs gets cancelled upon
integration over frequency $\omega'$ and only equal frequency
correlations survive.  

\subsection{Derivation of radiated power}

The net power transfer from the particle to the environment is
obtained using Eqs.~\ref{Pexp},\ref{Green},\ref{pfdt},\ref{Efdt}. The
power exchanged with the vacuum part of the geometry corresponding to
the vacuum part of the Green's function [in Eq.~\ref{Green}] is given
below:
\begin{align}
P_{\text{vac}}(\omega) = \frac{\omega^3}{\pi
  c^3}\Im\{\alpha_0(\omega)\}[\Theta_{T_p}-\Theta_{T_e}]
\end{align}
The radiated power corresponding to the reflected part of the Green's
function in Eq.~\ref{Green}, henceforth abbreviated as
$G_{jk}^{\text{ref}}( \rv_1,\rv_2,\omega)$, and originating from the
particle dipole moment fluctuations [Eq.\ref{pfdt}] is:
\begin{align}
P_{\text{ref}}^{p}(\omega) &= \omega^3 \mu_0
\Im[p_j^{\text{fl}^*}(\rv_1,\omega)\bigg(G_{jk}^{\text{ref}}(
  \rv_1,\rv_2,\omega)p_k(\rv_2, \omega) \bigg)_{\rv_2 \rightarrow
    \rv_1}] \nonumber \\ &= \omega^3 \mu_0 \frac{\epsilon_0}{\omega}
\Theta_{T_p} \Im\{\alpha_0(\omega)\}
\Im[\text{Tr}(G_{jk}^{\text{ref}})] \nonumber \\ &=
\frac{\omega^2}{c^2}\Theta_{T_p} \Im\{\alpha_0\} \Im\bigg[ \int
  \frac{k_\parallel dk_\parallel d\phi}{(2\pi)^2} \frac{ie^{2ik_z
      d}}{2k_z}
       [r_{ss}+r_{pp}\bigg(\frac{2k_\parallel^2}{k_0^2}-1\bigg)]
       \bigg]
\end{align}
The radiated power corresponding to the reflected part of the Green's
function, and originating from the environment field fluctuations
[Eq.\ref{Efdt}] is:
\begin{align}
P_{\text{ref}}^{\text{env}}(\omega) &= \omega \Im[ \epsilon_0
  \alpha^*(\omega) \langle E_j^{\text{fl}^*}E_j^{\text{fl}} \rangle ]
\nonumber \\ &= \omega\epsilon_0 \mu_0\omega\Theta_{T_e} \Im[
  \alpha^*(\omega) \Im[\text{Tr}(G_{jk}^{\text{ref}})]] \nonumber
\\ &= -\frac{\omega^2}{c^2}\Theta_{T_e} \Im\{\alpha_0\} \Im\bigg[ \int
  \frac{k_\parallel dk_\parallel d\phi}{(2\pi)^2} \frac{ie^{2ik_z
      d}}{2k_z}
       [r_{ss}+r_{pp}\bigg(\frac{2k_\parallel^2}{k_0^2}-1\bigg)]
       \bigg]
\end{align}
The spectrum of the net radiated power from the particle to the
environment provided in the main text is obtained by adding together
the above expressions.

\subsection{Derivation of force}

Because of the rotational symmetry of the spherical particle described
by isotropic polarizability, the direct interaction of the particle
with vacuum does not lead to any torque or force on the
particle. Therefore, we focus on the force and torque arising from the
reflected part of the Green's function. First, we calculate the force
spectrum [Eq.\ref{Fexp}] along $\mathbf{x}$ (parallel to the slab
surface) and $\mathbf{z}$ direction (perpendicular to the slab
surface). The force originating from the environment field
fluctuations is:
\begin{align}
F_x^{\text{env}}(\omega) &= \Re[\langle \epsilon_0\alpha_0
  E_j^{\text{fl}}\partial_x E_j^{\text{fl}^*} \rangle] = \int
\frac{d\omega}{2\pi} \Re[\langle \epsilon_0\alpha_0
  \underbrace{E_j^{\text{fl}}(\rv_1)\partial_{x_2}}_{\text{commuting
      terms}} E_j^{\text{fl}^*}(\rv_2) \rangle]_{\rv_2\rightarrow
  \rv_1} \nonumber \\ &= \Re[\langle \epsilon_0\alpha_0 \partial_{x_2}
  E_j^{\text{fl}}(\rv_1) E_j^{\text{fl}^*}(\rv_2)
  \rangle]_{\rv_2\rightarrow \rv_1} \nonumber \\ &= \epsilon_0
\Re[\alpha_0 \partial_{x_2} \langle
  E_j^{\text{fl}}(\rv_1)E_j^{\text{fl}^*}(\rv_2)\rangle]_{\rv_2\rightarrow
  \rv_1} \nonumber \\ &= \epsilon_0 \mu_0\omega\Theta_{T_e}
\Re[\alpha_0
  \frac{(-ik_x)G_{jj}(\rv_1,\rv_2)-(-ik_x)G_{jj}^*(\rv_2,\rv_1)}{2i}]_{\rv_2
  \rightarrow \rv_1} \nonumber \\ &= \frac{\omega}{c^2}\Theta_{T_e}
\Re[-i\alpha_0 \Im\bigg[ \text{Tr}\bigg[ \int \frac{d^2
        \kv_\parallel}{(2\pi)^2} \frac{ik_x}{2k_z} e^{2ik_z d}
      [(r_{ss}\ev_{s+}+r_{ps}\ev_{p+})\ev_{s-}^T +
        (r_{sp}\ev_{s+}+r_{pp}\ev_{p+})\ev_{p-}^T] \bigg] \bigg]]
\nonumber \\ &= \frac{\omega}{c^2}\Theta_{T_e}\Im\{\alpha_0\}
\Im\bigg[ \int \frac{k_\parallel dk_\parallel d\phi}{(2\pi)^2}
  \frac{ik_\parallel e^{2i k_z d}}{2k_z} \cos\phi [r_{ss} +
    (\frac{2k_\parallel^2}{k_0^2}-1)r_{pp}] \bigg]
\end{align}
Note that for reciprocal media, $r_{ss,pp}(\phi)=r_{ss,pp}(\phi+\pi)$
because of the time-reversal symmetry. It then follows that upon
integration over the angle $\phi$, the integrand at $\phi$ cancels
with the integrand at $\phi+\pi$ because of $\cos\phi$ term, leading
to zero parallel force. For nonreciprocal medium, parallel force is
not necessairly zero.

Similarly, the $\ev_z$ component of the force can be
calculated. Instead of $(-ik_x)$ terms in the above derivation, we
will get $(+ik_z)$ from the Green's function, eventually leading to
the following expression:
\begin{align}
F_z^{\text{env}}(\omega) =
\frac{-\omega}{c^2}\Theta_{T_e}\Im\{\alpha_0\} \Im\bigg[ \int
  \frac{k_\parallel dk_\parallel d\phi}{(2\pi)^2} \frac{i e^{2i k_z
      d}}{2} [r_{ss} + (\frac{2k_\parallel^2}{k_0^2}-1)r_{pp}] \bigg]
\end{align}
Note that for reciprocal media, no cancelleation over angular
integration occurs and the above expression leads to a force
perpendicular to the surface for both reciprocal and nonreciprocal
media. 

We will now calculate the force originating from the particle dipole
moment fluctuations:
\begin{align}
F_x^{p}(\omega) &= \Re[\langle p_j^{\text{fl}}\partial_x
  E_j^{\text{ind}^*}\rangle] = \int\frac{d\omega}{2\pi} \Re[\langle
  p_j^{\text{fl}}(\rv_1) \partial_{x_2} (\underbrace{\omega^2\mu_0
    G_{jk}(\rv_2,\rv_1)p_k^{\text{fl}}(\rv_1)}_{E_j^{\text{ind}}(\rv_2)})^*
  \rangle]_{\rv_2 \rightarrow \rv_1} \nonumber \\ &= \omega^2\mu_0
\Re[ \langle p_j^{\text{fl}}(\rv_1)p_k^{\text{fl}^*}(\rv_1) \rangle
  \bigg(\partial_{x_2} G_{jk}^*(\rv_2,\rv_1)\bigg)_{\rv_2\rightarrow
    \rv_1}] \nonumber \\ &= \omega^2\mu_0 \Re[
  \frac{\epsilon_0}{\omega}\Theta_{T_p} \Im\{\alpha_0\} \delta_{jk}
  \bigg( (-ik_x) G_{jk}^*(\rv_2,\rv_1)\bigg)_{\rv_2\rightarrow \rv_1}
] \nonumber \\ &= \frac{\omega}{c^2}\Theta_{T_p} \Im\{\alpha_0\}
\Im\bigg[ k_x G_{jj}^*(\rv_1,\rv_1) \bigg] \nonumber \\ &=
\frac{-\omega}{c^2}\Theta_{T_p}\Im\{\alpha_0\} \Im\bigg[ \int
  \frac{k_\parallel dk_\parallel d\phi}{(2\pi)^2} \frac{ik_\parallel
    e^{2i k_z d}}{2k_z} \cos\phi [r_{ss} +
    (\frac{2k_\parallel^2}{k_0^2}-1)r_{pp}] \bigg]
\end{align}
For the calculation of force along $\ev_z$ direction from particle
dipole moment fluctuations, the partial derivative $\partial_z$ leads
to $(-ik_z)$ instead of $+ik_z$ (for environment contribution) because
of the complex conjugation of the Green's function
i.e. $G_{jj}^*(\rv_1,\rv_1)$. The final expression for the force is:
\begin{align}
F_z^{\text{p}}(\omega) =
\frac{-\omega}{c^2}\Theta_{T_p}\Im\{\alpha_0\} \Im\bigg[ \int
  \frac{k_\parallel dk_\parallel d\phi}{(2\pi)^2} \frac{i e^{2i k_z
      d}}{2} [r_{ss} + (\frac{2k_\parallel^2}{k_0^2}-1)r_{pp}] \bigg]
\end{align}

The net parallel and perpendicular fluctuations-induced forces acting
on the particle are:
\begin{align}
F_x(\omega) &=
\frac{\omega}{c^2}[\Theta_{T_e}-\Theta_{T_p}]\Im\{\alpha_0\} \Im\bigg[
  \int \frac{k_\parallel dk_\parallel d\phi}{(2\pi)^2}
  \frac{ik_\parallel e^{2i k_z d}}{2k_z} \cos\phi [r_{ss} +
    (\frac{2k_\parallel^2}{k_0^2}-1)r_{pp}] \bigg] \\ F_y(\omega) &=
F_x(\omega,\cos\phi \rightarrow \sin\phi) \\ F_z(\omega) &=
\frac{-\omega}{c^2}[\Theta_{T_e}+\Theta_{T_p}]\Im\{\alpha_0\}
\Im\bigg[ \int \frac{k_\parallel dk_\parallel d\phi}{(2\pi)^2} \frac{i
    e^{2i k_z d}}{2} [r_{ss} + (\frac{2k_\parallel^2}{k_0^2}-1)r_{pp}]
  \bigg]
\end{align}

\subsection{Derivation of torque}

Similar to the calculation of force, we calculate the torque using the
reflected part of Green's function (indicating photons reflected from
the surface). The torque $M_x^{\text{env}}$ parallel to the surface
and originating from the environment field fluctuations is:
\begin{align}
M_x^{\text{env}}(\omega) &= \epsilon_0 \Re[\alpha_0 \langle (E_y E_z^*
  - E_z E_y^*) \rangle] = \int \frac{d\omega}{2\pi} \epsilon_0
\Re[\alpha_0 \times 2i \Im\{E_y E_z^*\} ] \nonumber \\ &=
-2\epsilon_0\mu_0\omega\Theta_{T_e}\Im\{\alpha_0\}
\Im\bigg[\frac{G_{yz}-G_{zy}^*}{2i} \bigg] \nonumber \\ &=
-\frac{\omega}{c^2}\Theta_{T_e}\Im\{\alpha_0\} \int \frac{d^2
  \kv_\parallel}{(2\pi)^2}\Im\bigg[
  \frac{-k_\parallel}{2k_0}\cos\phi\bigg( r_{sp}\frac{e^{2ik_z
      d}}{k_z} + r_{ps}^* \frac{e^{-2ik_z^* d}}{k_z^*}\bigg) -
  \frac{ik_\parallel}{k_0}\sin\phi \Im\bigg(r_{pp} \frac{e^{2ik_z
      d}}{k_0} \bigg)\bigg] \nonumber \\ &=
-\frac{\omega}{c^2}\Theta_{T_e}\Im\{\alpha_0\} \int \frac{k_\parallel
  dk_\parallel d\phi}{(2\pi)^2} \bigg[
  -\frac{k_\parallel}{2k_0}\cos\phi \Im\bigg(
  (r_{sp}-r_{ps})\frac{e^{2ik_z d}}{k_z}\bigg) -
  \frac{k_\parallel}{k_0^2}\sin\phi \Im\{r_{pp}e^{2ik_z d}\} \bigg]
\end{align}
The torque in the direction perpendicular to the surface is:
\begin{align}
M_z^{\text{env}}(\omega) &= \Re[\alpha_0 \times 2i \Im\{ E_x E_y^*\} ]
\nonumber \\ &= -2\epsilon_0\mu_0\omega\Theta_{T_e} \Im\bigg[
  \frac{G_{xy}-G_{yx}^*}{2i}\bigg] \nonumber \\ &=
-\frac{\omega}{c^2}\Theta_{T_e} \Im\{\alpha_0\} \int \frac{k_\parallel
  dk_\parallel d\phi}{(2\pi)^2}
\frac{1}{2k_0}\Im[(r_{sp}+r_{ps})e^{2ik_z d}]
\end{align}

The torque components $M_{x,z}^{\text{p}}$ due to photons generated by
particle dipole moment fluctuations are:
\begin{align}
M_x^{\text{p}}(\omega) &= \omega^2 \mu_0 \Re[\langle
  \underbrace{p_y^{\text{fl}}(G_{zy}p_y^{\text{fl}})^* -
    p_z^{\text{fl}}(G_{yz}p_z^{\text{fl}})^*}_{\text{taking into
      account the correlations between dipole moments}} \rangle ]
\nonumber \\ &= \omega^2 \mu_0
\frac{\epsilon_0}{\omega}\Theta_{T_p}\Im\{\alpha_0\} \Re[ G_{zy}^* -
  G_{yz}^* ] \nonumber \\ &= \frac{\omega}{c^2}\Theta_{T_p}
\Im\{\alpha_0\} \Im[iG_{zy}^*-iG_{yz}^*] \nonumber \\ &=
2\frac{\omega}{c^2}\Theta_{T_p} \Im\{\alpha_0\}
\Im[\frac{G_{yz}-G_{zy}^*}{2i}] \nonumber \\ &=
\frac{\omega}{c^2}\Theta_{T_p}\Im\{\alpha_0\} \int \frac{k_\parallel
  dk_\parallel d\phi}{(2\pi)^2} \bigg[
  -\frac{k_\parallel}{2k_0}\cos\phi \Im\bigg(
  (r_{sp}-r_{ps})\frac{e^{2ik_z d}}{k_z}\bigg) -
  \frac{k_\parallel}{k_0^2}\sin\phi \Im\{r_{pp}e^{2ik_z d}\} \bigg]
\end{align}
The torque along $\mathbf{z}$ direction is:
\begin{align}
M_z^{\text{p}}(\omega) &= 2\frac{\omega}{c^2}\Theta_{T_p}
\Im\{\alpha_0\} \Im[\frac{G_{xy}-G_{yx}^*}{2i}] \nonumber \\ &=
\frac{\omega}{c^2}\Theta_{T_p} \Im\{\alpha_0\} \int \frac{k_\parallel
  dk_\parallel d\phi}{(2\pi)^2}
\frac{1}{2k_0}\Im[(r_{sp}+r_{ps})e^{2ik_z d}]
\end{align}
The final expressions for the fluctuations-induced torque spectrum are:
\begin{align}
M_x(\omega) &=
\frac{\omega}{c^2}(\Theta_{T_e}-\Theta_{T_p})\Im\{\alpha_0\} \int
\frac{k_\parallel dk_\parallel d\phi}{(2\pi)^2} \bigg[
  \frac{k_\parallel}{2k_0}\cos\phi \Im\bigg(
  (r_{sp}-r_{ps})\frac{e^{2ik_z d}}{k_z}\bigg) +
  \frac{k_\parallel}{k_0^2}\sin\phi \Im\{r_{pp}e^{2ik_z d}\} \bigg]
\\ M_y(\omega) &=
\frac{\omega}{c^2}(\Theta_{T_e}-\Theta_{T_p})\Im\{\alpha_0\} \int
\frac{k_\parallel dk_\parallel d\phi}{(2\pi)^2} \bigg[
  \frac{k_\parallel}{2k_0}\sin\phi \Im\bigg(
  (r_{sp}-r_{ps})\frac{e^{2ik_z d}}{k_z}\bigg) -
  \frac{k_\parallel}{k_0^2}\cos\phi \Im\{r_{pp}e^{2ik_z d}\} \bigg]
\\ M_z(\omega) &= -\frac{\omega}{c^2}(\Theta_{T_e}-\Theta_{T_p})
\Im\{\alpha_0\} \int \frac{k_\parallel dk_\parallel d\phi}{(2\pi)^2}
\frac{1}{2k_0}\Im[(r_{sp}+r_{ps})e^{2ik_z d}]
\end{align}
For reciprocal media,
$r_{ss,pp}(\theta,\phi)=r_{ss,pp}(\theta,\phi+\pi)$ and
$r_{sp}(\theta,\phi)=-r_{ps}(\theta,\phi)$. By summing the integrands
at $\phi$ and $\phi+\pi$, it follows that particle does not experience
any torque. For InSb slab with magnetic field parallel to the surface,
this is no longer true and a lateral torque parallel to surface is
observed. However, the Fresnel coefficients for this system also
satisfy the condition $r_{sp}+r_{ps}=0$~\cite{khandekar2020new}
leading to a zero torque along $\mathbf{z}$ axis.

We note that all components of force and torque can be derived by
dividing the integrand in the power spectrum by energy
$\hbar\omega$, multiplying by linear momentum vector
$\hbar(\kp\cos\phi,\kp\sin\phi,kz)$ and spin angular momentum vector
$\hbar(\sin\phi,-\cos\phi,0)$ of polaritons, and accounting for
whether the photon is emitted or absorbed by the particle. This
physically meaningful alternative derivation is exact for force
calculation which is evident from the final expressions of the spectra
above. It is approximate for torque because the spin angular momentum
$\hbar(\sin\phi,-\cos\phi,0)$ is exact only for large wavevector
polaritons ($\kp \gg k_0$). This is proved in the following.  We can
calculate the angular momentum transferred for each photon exchanged
which is characterized by $(\omega,\kp,\phi)$:
\begin{align}
M_{x}(\omega,\kp,\phi)= \frac{\bigg[ \frac{k_\parallel}{2k_0}\cos\phi
    \Im\bigg( (r_{sp}-r_{ps})\frac{e^{2ik_z d}}{k_z}\bigg) +
    \frac{k_\parallel}{k_0^2}\sin\phi \Im\{r_{pp}e^{2ik_z d}\}
    \bigg]}{\Im\bigg[\frac{ie^{2ik_z d}}{2k_z}
    [r_{ss}+r_{pp}\bigg(\frac{2k_\parallel^2}{k_0^2}-1\bigg)\bigg]}\hbar
  \\ M_{y}(\omega,\kp,\phi)= \frac{\bigg[
      \frac{k_\parallel}{2k_0}\sin\phi \Im\bigg(
      (r_{sp}-r_{ps})\frac{e^{2ik_z d}}{k_z}\bigg) -
      \frac{k_\parallel}{k_0^2}\cos\phi \Im\{r_{pp}e^{2ik_z d}\}
      \bigg]}{\Im\bigg[\frac{ie^{2ik_z d}}{2k_z}
      [r_{ss}+r_{pp}\bigg(\frac{2k_\parallel^2}{k_0^2}-1\bigg)\bigg]}\hbar
\end{align}

For evanescent waves ($\kp > k_0$), $k_z=i|k_z|$ is purely imaginary
which simplifies the above expressions:
\begin{align}
M_{x}(\omega,\kp,\phi)= \frac{\bigg[
    \frac{k_\parallel}{2k_0|k_z|}\cos\phi
    \Re\bigg(-(r_{sp}-r_{ps})\bigg) +
    \frac{k_\parallel}{k_0^2}\sin\phi \Im\{r_{pp}\} \bigg]}{\bigg[
    \frac{\Im(r_{ss})}{2|k_z|} +
    \frac{\Im(r_{pp})}{2|k_z|}\bigg(\frac{2k_\parallel^2}{k_0^2}-1\bigg)\bigg]}\hbar
\\ M_{y}(\omega,\kp,\phi)= \frac{\bigg[
    \frac{k_\parallel}{2k_0|k_z|}\sin\phi
    \Re\bigg(-(r_{sp}-r_{ps})\bigg) -
    \frac{k_\parallel}{k_0^2}\cos\phi \Im\{r_{pp}\} \bigg]}{\bigg[
    \frac{\Im(r_{ss})}{2|k_z|} +
    \frac{\Im(r_{pp})}{2|k_z|}\bigg(\frac{2k_\parallel^2}{k_0^2}-1\bigg)\bigg]}\hbar
\end{align}

For evanescent waves carrying large momentum $\kp \gg k_0$ which leads
to $|k_z|\approx \kp$, the above expressions further simplify to:
\begin{align}
M_x(\omega,\kp\gg k_0,\phi)= (\sin\phi) \hbar, \hspace{40pt}
M_y(\omega,\kp \gg k_0,\phi)= (-\cos\phi) \hbar
\end{align}

\section{Rotational dynamics of the particle in imperfect vacuum}

We consider a potential experiment where a particle is levitated in
the near-field of doped InSb plate inside a vacuum chamber and the
magnetic field along $\mathbf{x}$ axis is switched on at time
$t=0$. The particle at temperature $T_p \neq T_e$ experiences a
lateral Casimir torque $M_x$ which causes an angular acceleration
about $\mathbf{x}$ axis. As the particle's angular velocity ($\Omega$)
increases, there is also an additional damping torque from the
residual air molecules inside the vacuum chamber. The damping torque
opposes the angular velocity and it can be described
as~\cite{xu2020enhancement,ahn2018optically}:
\begin{align}
M_{\text{damp}} = -\frac{\Omega p_{\text{gas}} \pi
  (2R)^4}{11.976}\sqrt{\frac{2m_{\text{gas}}}{k_B T_e}} = -
\gamma_{\text{damp}}\Omega
\end{align}
where $R$ is the radius of the nanoparticle, $m_{\text{gas}}=4.8\times
10^{-26}$kg is the average mass of air molecule, and $p_{\text{gas}}$
is the air pressure. Using the particle's moment of inertia
$I_x=\frac{2}{5}m_{\text{p}}R^2$, the evolution of its angular
velocity follows the following stochastic equation derived using
Brownian theory~\cite{higham2001algorithmic}:
\begin{align}
 d\Omega = \frac{M_x}{I_x}dt -
 \frac{\gamma_{\text{damp}}}{I_x}\Omega dt +
 \sqrt{\frac{2\gamma_{\text{damp}}k_BT_e}{I_x^2}} dW
\end{align}  
where $dW$ is a normally distributed random number of mean $\langle dW
\rangle = 0$ and standard deviation $\langle dW^2 \rangle = dt$ and
$dt$ is the time step much smaller than the relaxation timescale
$\mathcal{O}(I_x/\gamma_{\text{damp}})$. The term multiplying $dW$,
also known as fluctuation--dissipation theorem (FDT), indicates that
the strength of the fluctuations is related to the dissipation
rate. The above equation is simulated using Euler Maruyama method over
a large number of trajectories~\cite{higham2001algorithmic}. In the
absence of gyrotropy-induced lateral torque ($M_x=0$), the specific
form of FDT ensures that the average rotational energy of the particle
given by $\frac{1}{2}I_x\Omega^2$ is equal to $k_B T_e/2$ at
equilibrium with the environment at temperature $T_e$ (equipartition
law). At equilibrium, the average angular velocity $\langle \Omega
\rangle =0$ and the standard deviation $\sqrt{\langle \Omega^2
  \rangle}=k_B T_e/ I_x$. For AgBr and NaCl nanoparticles in an
environment at $T_e=300$K (considered in the main text), the typical
standard deviation is $\sqrt{\langle \Omega^2 \rangle} \approx
10^4$rad/s.

In presence of lateral nonequilibrium Casimir torque ($M_x\neq 0$),
the steady-state angular velocity is
$\Omega_{\text{ss}}=M_x/\gamma_{\text{damp}}$ ignoring the fluctuating
noise. It is straightforward to numerically verify that the fluctating
noise can be ignored if $\Omega_{\text{ss}} \gg 10^4$rad/s. For
realistic vacuum chambers of gas pressure $10^{-5}$torr and lateral
nonequilibrium torque $M_x \gtrsim 10^{-27}$Nm attained in the
near-field of InSb slab, we find $\Omega_{\text{ss}} \gg
10^4$rad/s. For smaller values of torque away from the slab ($M_x <
10^{-27}$Nm), in general, full numerical simulations including the
fluctuating thermal noise are required for calculating the
steady-state mean angular velocity. For predictions related to
experiments involving levitated particles, both shot noise of the
laser and the additional thermal noise can be considered in the above
equation to obtain the steady-state mean angular velocities.

{\bf Time taken to reach steady state}: Figure 3 in the main text
reveals the rotational steady state angular velocities
$\Omega_{\text{ss}} \gtrsim 10^6$rad/s for AgBr and NaCl particles of
radius $R=200$nm in a vacuum chamber of
$p_{\text{gas}}=10^{-5}$torr. For the above parameters, the time
constant $I_x/\gamma_{\text{damp}}$ is of the order of minutes for
AgBr ($\sim 140$s.) and NaCl ($\sim 48$s.)  particles. As the magnetic
field of $1T$ is turned on at $t=0$, the particles will reach these
steady state rotational velocities in few minutes.

{\bf Angular velocity dependence of the torque}: We note that, in the
above equation, the lateral Casimir torque $M_x$ is derived for a
non-rotating particle. The effect of the particle's rotational
velocity $\Omega$ on the magnitude of this torque is negligible as
long as its rotational velocity is much smaller than the emission
frequencies ($\Omega \ll \omega$). For our situation, $\omega \sim
10^{13}$rad/s while $\Omega_{ss} \sim 10^7$rad/s and hence ignoring
the dependence of $M_x$ on $\Omega$ is justified. Nonetheless, we
derive $M_x(\Omega)$ below and prove this point. We follow the
derivation similar to that of ref.~\cite{Manjavacas2017lateral} for
the calculation of torque on rotating particles. The torque
originating from the particle's dipole moment fluctuations is
calculated in the laboratory frame. For a particle rotating at angular
velocity $\Omega$ about $\mathbf{x}$ axis, the fluctuating dipole
moments in the lab frame $p_j^l$ are related to the fluctuating dipole
moments in the rotating frame $p_j^{\text{fl}}$ (FDT is well-defined
in the rotating frame~\cite{Manjavacas2017lateral}) as:
\begin{align*}
p_y^l &= \frac{1}{2}[ p_{y,\omega}^{\text{fl}}e^{-i\omega_- t} +
  p_{y,\omega}^{\text{fl}}e^{-i\omega_+ t}+ p_{y,\omega}^{\text{fl}^*}
  e^{i\omega_+ t} + p_{y,\omega}^{\text{fl}^*} e^{i\omega_- t} + i
  p_{z,\omega}^{\text{fl}}e^{-i\omega_- t} - i
  p_{z,\omega}^{\text{fl}}e^{-i\omega_+ t} + i
  p_{z,\omega}^{\text{fl}^*}e^{i\omega_+ t} - i
  p_{z,\omega}^{\text{fl}^*}e^{i\omega_- t}] \\ p_z^l &=
\frac{1}{2}[-i p_{y,\omega}^{\text{fl}}e^{-i\omega_- t} + i
  p_{y,\omega}^{\text{fl}}e^{-i\omega_+ t}
  -ip_{y,\omega}^{\text{fl}^*}e^{i\omega_+ t}
  +ip_{y,\omega}^{\text{fl}^*}e^{i\omega_- t} +
  p_{z,\omega}^{\text{fl}}e^{-i\omega_- t} +
  p_{z,\omega}^{\text{fl}}e^{-i\omega_+ t} +
  p_{z,\omega}^{\text{fl}^*}e^{i\omega_+ t} +
  p_{z,\omega}^{\text{fl}^*}e^{i\omega_- t}] \\ p_x^l &=
p_{x,\omega}^{\text{fl}}e^{-i\omega t} +
p_{x,\omega}^{\text{fl}^*}e^{i\omega t}
\end{align*}
where $\omega_{\pm} = \omega \pm \Omega$ and both positive and
negative frequencies are considered together such that the final
integration is performed only over positive frequencies
($\int_0^{\infty} d\omega/2\pi$). The fluctuating dipole moments
$p_{j,\omega}^{\text{fl}}$ satisfy the FDTs above and are used to
calculate the resulting torque. This calculation also requires induced
electric fields which are obtained using the Green's function
[$E_{j}^{\text{ind}}=G_{jk,\omega}p_{k,\omega_m}^{\text{fl}}e^{-i\omega_m
    t}$ for $\omega_m \in [\omega,\omega_+,\omega_-]$ and its complex
  conjugate].

The calculation of torque originating from the electric field
fluctuations requires the electric fields in the rotating frame of the
particle. For the particle rotating at angular velocity $\Omega$, the
field fluctuations $E_j^r$ in the rotating frame are:
\begin{align*}
E_{y}^r &= \frac{1}{2} [E_{y,\omega}^{\text{fl}} e^{-i\omega_- t} +
  E_{y,\omega}^{\text{fl}} e^{-i\omega_+ t} +
  E_{y,\omega}^{\text{fl}^*}e^{i\omega_+ t} +
  E_{y,\omega}^{\text{fl}^*} e^{i\omega_- t} - i
  E_{z,\omega}e^{-i\omega_- t} + iE_{z,\omega}e^{-i\omega_+ t} -
  iE_{z,\omega}^{\text{fl}^*}e^{i\omega_+ t} +
  iE_{z,\omega}^{\text{fl}^*}e^{i\omega_- t} ] \\ E_{z}^r &=
\frac{1}{2} [iE_{y,\omega}^{\text{fl}}e^{-i\omega_- t} -
  iE_{y,\omega}^{\text{fl}} e^{-i\omega_+ t} +
  iE_{y,\omega}^{\text{fl}^*} e^{i\omega_+ t} -
  iE_{y,\omega}^{\text{fl}^*}e^{i\omega_- t}+ E_{z,\omega}^{\text{fl}}
  e^{-i\omega_- t} + E_{z,\omega}^{\text{fl}} e^{-i\omega_+ t} +
  E_{z,\omega}^{\text{fl}^*}e^{i\omega_+ t} +
  E_{z,\omega}^{\text{fl}^*} e^{i\omega_- t} ] \\ E_{x}^r &=
E_{x,\omega}^{\text{fl}}e^{-i\omega t} +
E_{x,\omega}^{\text{fl}^*}e^{i\omega t}
\end{align*}
Using the induced dipole moments in the rotating frame,
$p_j^{r,\text{ind}}=\epsilon_0\alpha_0 E_{j}^r$, the torque due to
electric field fluctuations can be obtained. After some
straightforward algebra, we get the following torque due to the
reflected part of the Green's function:
\begin{align}
  M_x^{e} &= \frac{\epsilon_0 \Theta_{T_e}}{\omega} \bigg[
    \Im(\alpha_{\omega_-}+\alpha_{\omega_+})\Re(G^t_{yz,\omega}-G^t_{zy,\omega})
    + \Im(G^t_{yy}+G^t_{zz})\Im(\alpha_{\omega_-}-\alpha_{\omega_+})
    \bigg] \\ M_x^{p} &= \frac{\epsilon_0\Theta_{T_p}}{\omega}
  \Im(\alpha_{\omega})\bigg[ \Im(G^t_{zz,\omega_-}-G^t_{zz,\omega_+})+
    \Im(G^t_{yy,\omega_-}-G^t_{yy,\omega_+}) -
    \Re(G^t_{yz,\omega_+}-G^t_{zy,\omega_+}) -
    \Re(G^t_{yz,\omega_-}-G^t_{zy,\omega_-}) \bigg]
\end{align}
where $G^t = \omega^2\mu_0 G$ is used for simplifying the
expressions. The subscripts indicate the frequencies at which the
related terms are to be evaluated. The required expressions for $G^t$
are:
\begin{align*}
G^t_{yz,\omega}-G^t_{zy,\omega}&=\omega^2\mu_0 \int \frac{\kp d\kp
  d\phi}{(2\pi)^2} \frac{ie^{2i k_z d}}{2k_z}
[-(r_{sp}-r_{ps})\frac{\kp}{k_0}\cos\phi - 2r_{pp}\frac{\kp
    k_z}{k_0^2}] \\ G^t_{yy,\omega}&= \omega^2\mu_0 \int \frac{\kp d\kp
  d\phi}{(2\pi)^2} \frac{ie^{2i k_z d}}{2k_z} [r_{ss}\cos^2\phi +
  (r_{ps}-r_{sp})\frac{k_z}{k_0}\sin\phi \cos\phi -
  r_{pp}\frac{k_z^2}{k_0^2}\sin^2\phi] \\ G^t_{zz,\omega} &=
\omega^2\mu_0 \int \frac{\kp d\kp d\phi}{(2\pi)^2} \frac{ie^{2i k_z
    d}}{2k_z} r_{pp}\frac{\kp^2}{k_0^2}
\end{align*}
The torque because of the vacuum part of the Green's function in
Eq.~\ref{Green} (indicating direct interaction with vacuum half-space)
is:
\begin{align}
M_x^{\text{e,vac}} &= \frac{\epsilon_0\Theta_{T_e}}{\omega}
\Im(\alpha_{\omega_-}-\alpha_{\omega_+})
\frac{2\omega^3}{3\pi\epsilon_0 c^3} = \frac{2\omega^2}{3\pi
  c^3}\Im(\alpha_{\omega_-}-\alpha_{\omega_+})\Theta_{T_e}
\\ M_x^{\text{p,vac}} &=
\frac{\epsilon_0\Theta_{T_p}}{\omega}\Im(\alpha_{\omega}) \frac{2
  (\omega_-^3-\omega_+^3)}{3\pi \epsilon_0 c^3} =
\frac{2(\omega_-^3-\omega_+^3)}{3\pi \omega c^3}
\Im(\alpha_\omega)\Theta_{T_p}
\end{align}
In the following, we use the above expressions to compute the lateral
Casimir torque for the case of AgBr and NaCl particles considered in
the main text.

\begin{figure}[t!]
  \centering\includegraphics[width=0.4\linewidth]{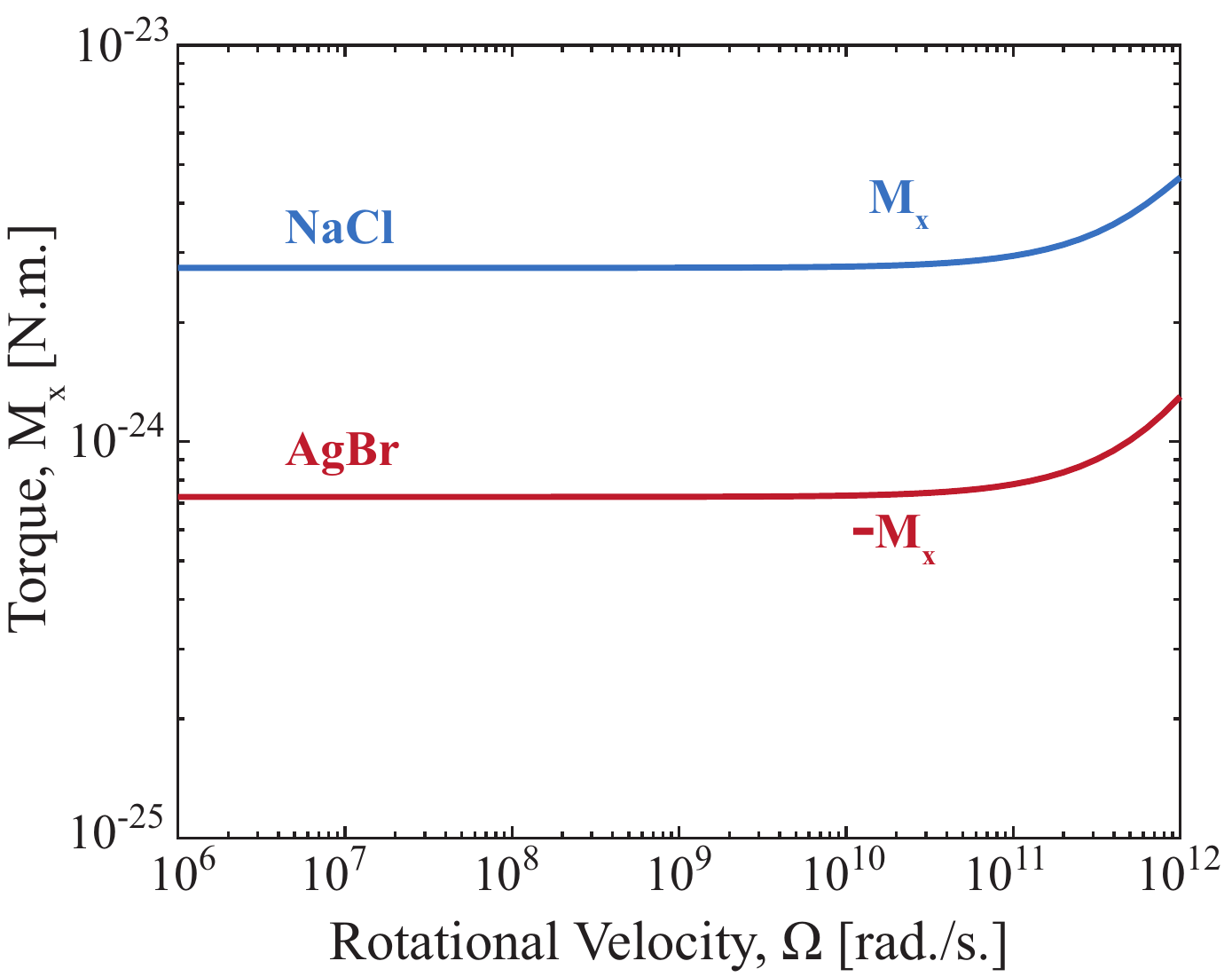}
  \caption{Figure shows the angular velocity dependence of the lateral
    nonequilibrium torque on AgBr (red) and NaCl (blue) nanoparticle
    of $R=200$nm at a distance $d_s=100$nm from the surface of InSb
    half-space in $B=1T$ magnetic field along $\mathbf{x}$. The
    operating temperatures are $T_p=400$K and $T_e=300$K. The
    magnitude of the torque is affected appreciably at large angular
    velocities of $\Omega \sim \mathcal{O}(10^{11})$rad/s because the
    rotational Doppler effect on thermal emission dominant at THz
    frequencies ($\omega \gtrsim 10^{12}$rad/s) is non-negligible.}
  \label{torque}
\end{figure} 

As explained in the main text, we focus on two cases of AgBr and NaCl
nanoparticles of radius $R=200$nm in the near-field of doped InSb slab
in magnetic field $B=1$T along $\mathbf{x}$ axis of the geometry. The
particles experience oppositely directed lateral nonequilibrium
Casimir torque and hence rotate with opposite angular
velocities. Figure~\ref{torque} demonstrates the dependence of the
fluctuations-induced torque on the angular velocity of the
particle. For this demonstration, we assumed both particles to be at
surface-to-surface distance $d_s=0.1\mu$m. from the slab surface and
at temperature $T_p=400$K. The temperature of the surrounding
environment is $T_e=300$K. Intuitively, it follows from the above
torque expressions containing terms such as $\omega_{\pm}=\omega \pm
\Omega$ that the angular velocity will alter the magnitude of the
torque only when it is comparable to the emission frequencies which
lie in the THz range. This is evident from Fig.\ref{torque} where the
torque is constant for rotation velocities less than $\Omega =
10^{10}$rad/s. For particles reaching the rotation speeds beyond these
values, we should use the above angular-velocity-dependent torque
expressions for calculations. However, since the damping torque of the
imperfect vacuum chamber balances the lateral Casimir torque at
smaller rotation speeds in the MHz to GHz range, it suffices to use
the torque expressions derived for steady particles. We also note that
at rotation speeds beyond 10GHz where the effect of the rotation speed
on the magnitude of the torque is noticeable, the particle may
disintegrate because of the centrifugal stress. These effects depend
on the ultimate tensile strength of the material, and must be taken
into account at these high rotational
velocities~\cite{reimann2018rotation,ahn2018optically}.

\bibliography{photon}